\documentclass[%
reprint,
superscriptaddress,
nofootinbib,
 amsmath,amssymb,
 aps,
]{revtex4-2}

\usepackage{graphicx}
\usepackage{bm}
\usepackage{hyperref}
\usepackage{color}
\usepackage{amsmath,amsthm}
\usepackage{comment}

\newtheorem*{conjecture*}{Conjecture}

\renewcommand{\d}{\mathrm d}

\let\Re\relax \let\Im\relax
\DeclareMathOperator{\Re}{Re}
\DeclareMathOperator{\Im}{Im}
\DeclareMathOperator{\Tr}{Tr}
\DeclareMathOperator{\Erf}{Erf}

\DeclareMathOperator{\sgn}{sgn}

\begin{document}

\title{Normalizing Flows and the Real-Time Sign Problem}

\author{Scott Lawrence}
\email{scott.lawrence-1@colorado.edu}
\affiliation{Department of Physics, University of Colorado, Boulder, CO 80309, USA}

\author{Yukari Yamauchi}
\email{yyukari@umd.edu}
\affiliation{Department of Physics, University of Maryland, College Park, Maryland 20742, USA}

\begin{abstract}
Normalizing flows have recently been applied to the problem of accelerating Markov chains in lattice field theory. We propose a generalization of normalizing flows that allows them to applied to theories with a sign problem. These complex normalizing flows are closely related to contour deformations (i.e.\ the generalized Lefschetz thimble method), which been applied to sign problems in the past. We discuss the question of the existence of normalizing flows: they do not exist in the most general case, but we argue that exact normalizing flows are likely to exist for many physically interesting problems, including cases where the Lefschetz thimble decomposition has an intractable sign problem. Finally, normalizing flows can be constructed in perturbation theory. We give numerical results on their effectiveness across a range of couplings for the Schwinger-Keldysh sign problem associated to a real scalar field in $0+1$ dimensions.
\end{abstract}

\maketitle

\section{Introduction}\label{sec:intro}
Monte Carlo methods, applied to lattice quantum field theory, are unique in providing nonperturbative access to observables in QCD and other field theories. These methods are not, however, equally applicable to all theories and observables. In particular, when applied to theories with a finite density of relativistic fermions, or to observables involving real-time evolution, lattice Monte Carlo methods are afflicted by the so-called \emph{sign problem}. This obstacle to computing quantum real-time dynamics has persisted for a considerable time, and is a central motivation for the use of quantum computers in high energy and nuclear physics.

Lattice methods work by framing the observable to be computed as a ratio of high-dimensional integrals. Spacetime is discretized, and Feynman's path integral becomes a finite- (but large-) dimensional integral. For many theories, this procedure results in a probability distribution over field configurations, which can be importance sampled with Markov chain Monte Carlo methods. However, in the case of a finite density of relativistic fermions, or nonequilibirium calculations, the Boltzmann factor $e^{-S}$ is generally complex, and cannot be treated as a probability distribution. In such cases, a standard approach is to sample according to the ``quenched'' Boltzmann factor $e^{-\Re S}$, and include the phases by reweighting. The cost of reweighting is generically exponential in the spacetime volume of the system being simulated.

Recent work has introduced \emph{normalizing flows} as a tool for accelerating Markov chain Monte Carlo methods~\cite{Albergo:2019eim,Kanwar:2020xzo,Boyda:2020hsi,Nicoli:2020evf,Nicoli:2020njz}. The idea is to construct (usually training by gradient descent) a generative model that samples approximately according to the lattice Boltzmann factor. Either by reweighting or by using the model to create proposals for a Markov chain, the systemic bias of training is removed. This method is particularly anticipated to reduce the cost associated with the approach to the continuum limit (``critical slowing down'').

Normalizing flows, as usually formulated, are not directly helpful for the sign problem: a generative model necessarily models a real probability distribution, rather than the complex weights associated to lattice models with a sign problem. In this paper, we show how normalizing flows may be generalized to alleviate or remove a sign problem. The core of this idea is the observation that a normalizing flow, suitably generalized, implicitly defines an integration contour along which the sign problem may be alleviated or removed. These \emph{complex} normalizing flows are thus in the same family of methods as the Lefschetz thimble approach~\cite{Cristoforetti:2012su}, the generalized Lefschetz thimble method~\cite{Alexandru:2015sua,Alexandru:2016ejd}, and the search for sign-optimized manifolds~\cite{Mori:2017nwj,Mori:2017pne,Alexandru:2018ddf,Alexandru:2018fqp}.

Complex normalizing flows exist only when a manifold is available that exactly solves the sign problem. We discuss the conditions under which such perfect manifolds exist. By modifying the holomorphic gradient flow of~\cite{Alexandru:2015sua,Alexandru:2015xva}, we argue that \emph{locally} perfect manifolds, on which there are no local fluctuations of the phase of the Boltzmann factor, always exist. These manifolds may nonetheless possess a \emph{global} sign problem: different components of the manifold, separated by singularities of the action, may contribute with the integral with different phases, and therefore (partially) cancel. Conditioned on a mild conjecture regarding the dependence of locally perfect manifolds on the parameters of the action, we argue that globally perfect manifolds are likely to exist for a broad class of physical systems, including the Schwinger-Keldysh sign problem.

When a manifold is available that merely approximately solves the sign problem, an approximate normalizing flow exists. A simple physical argument suggests that for many problems of physical relevance, manifolds that approximately solve the sign problem (with the approximation getting \emph{better} in the infinite volume limit) should be available.

We also find that the tool of normalizing flows results in a method for perturbatively approximating sign-problem-ameliorating integration contours, as well as a new approach for machine learning of such contours (a prospect previously explored in~\cite{Alexandru:2017czx,Alexandru:2018fqp,Wynen:2020uzx,Mori:2017nwj,Mori:2017pne}). We explore the in-practice effectiveness of the perturbatively constructed flow with numerical experiments on modest $0+1$ lattices. This method does not appear to have scaling properties that would allow it to be used, at least without serious improvement, in higher-dimensional theories.

Finally, a perturbative view of normalizing flows gives rise to a method of computing lattice expectation values by solving a certain high-dimensional first-order partial differential equation. We demonstrate this method on lattice scalar field theory. Unfortunately, this is mostly a curiosity, chiefly because the practical algorithm for solving the differential equations represents an uncontrolled approximation. On theories with no sign problem, the fact that reliable error bars are not available renders it inferior to standard methods; on theories with a sign problem, solving the differential equation turns out to be hard in practice (for reasons apparently closely connected to the sign problem itself).

The remainder of this paper is structured as follows. In Sec.~\ref{sec:sk} we describe the lattice Schwinger-Keldysh formalism and the origin of the sign problem. Sec.~\ref{sec:flows-integrals} details the generalization of normalizing flows to the complex setting, and shows how they relate to contour integrals and the generalized thimble method. We discuss the question of the existence of complex normalizing flows (and correspondingly, manifolds that solve the sign problem) in Sec.~\ref{sec:existence}; the notions of ``global'' and ``local'' sign problems are defined here. Perturbative constructions of complex normalizing flows are given in Sec.~\ref{sec:perturbation}, with numerical experiments characterizing their effectiveness. Finally, Sec.~\ref{sec:discussion} outlines future avenues to explore.

\section{Lattice Schwinger-Keldysh}\label{sec:sk}
The lattice Schwinger-Keldysh method was introduced in~\cite{Alexandru:2016gsd,Alexandru:2017lqr}, for use with the generalized Lefschetz thimble method, as a formalism for computing real-time observables --- that is, observables where the operators have some time separation. The Schwinger-Keldysh action is readily derived by considering a lattice field theory in the Hamiltonian formulation. We are interested in a time-separated observable, of the form $\langle \mathcal O(t) \mathcal O(0)\rangle$, with the expectation value taken in a thermal ensemble of inverse temperature $\beta$. Removing time-dependences from the operators, this expectation value can be written
\begin{equation}
\langle \mathcal O(t) \mathcal O(0)\rangle
=
\frac{\Tr e^{-\beta H}e^{i H t} \mathcal O e^{-i H t} \mathcal O}{\Tr e^{-\beta H}}
\text.
\end{equation}
The ordinary lattice path integral involves only the imaginary-time operator, $e^{-\beta H}$. That operator is split up into a product of many $e^{-a_\tau H}$, each of which is Trotterized. Resolutions of the identity are inserted between each pair of operators, resulting in an integral over all (discrete) paths of field configurations.

The Schwinger-Keldysh path integral does not differ in its derivation. After all time-evolution operators (whether real or imaginary) are Trotterized and the field integrals inserted, the expectation value is given by
\begin{equation}
\langle \mathcal O(t) \mathcal O(0)\rangle
= \frac{\int \mathcal D\phi\; e^{-S[\phi]} \;\mathcal O(t) \mathcal O(0)}{\int \mathcal D\phi \;e^{-S[\phi]}}
\text,
\end{equation}
with the (Euclidean) action, in the case of a single real scalar field,
\begin{widetext}
\begin{equation}\label{eq:sk}
    S =
    \sum_{t,x}
    \frac{(\phi_{x,t} - \phi_{x,t+1})^2}{2 a_0(t)}
    +
    \sum_t
    \frac{a_0(t) + a_0(t-1)}{2}
    \left[
    \sum_{\langle x x'\rangle}
    \frac{(\phi_{x,t} - \phi_{x',t})^2}{2 a_x^2}
    +
    \sum_{x}
    \left(
    \frac{m^2}{2} \phi_{x,t}^2
+
\frac{\lambda}{4!} \phi_{x,t}^4
    \right)
\right]
\text.
\end{equation}
\end{widetext}
Here $m$ is the bare mass and $\lambda$ the coupling. Because some of the Trotterized time-evolution operators were imaginary time and others were real time, the timelike lattice spacing $a_0$ is taken to vary over the lattice. In this paper we will take it to be defined by the ``S-contour'', although other choices are possible:
\begin{equation}
    a_0(t) = \begin{cases}
    -i & t \in [0,N_t)\\
    1 & t \in [N_t, N_t + N_\beta / 2)\\
    i & t \in [N_t + N_\beta / 2,2 N_t + N_\beta / 2)\\
    1 & t \in [2 N_t + N_\beta / 2, 2 N_t + N_\beta)
    \end{cases}
    \text.
\end{equation}
Here $N_t$ and $N_\beta$ denote the number of real-time and thermodynamic time evolution steps, respectively. This choice of action is equivalent to an $O(a_0^2)$ Trotter approximation to $e^{-\beta H /2}e^{i H t} e^{-\beta H/2} e^{-i H t}$.

If there were no timeslices with $\Im a_0 \ne 0$, the action would be real. In that case, the Metropolis method gives an algorithm by which the computer can sample from the probability distribution proportional to $e^{-S}$. Expectation values with respect to that distribution correspond to physical expectation values.

As things are, the action is not pure real, and the Boltzmann factor does not correspond to any probability distribution --- at least, not any distribution of real-valued fields. The standard approach at this point is to sample with respect to the quenched Boltzmann factor $e^{-\Re S}$, and then reweight, computing observables as
\begin{equation}
    \langle \mathcal O\rangle
    =
    \frac{\langle e^{-i S_I} \mathcal O\rangle_Q}{\langle  e^{-i S_I}\rangle_Q}
\end{equation}
Here $\langle\cdot\rangle_Q$ denotes an expectation value with respect to the quenched distribution. The denominator, $\langle e^{-i S_I}\rangle_Q$, is known as the average phase, and characteristically decays exponentially in the spacetime volume of the system. A simple but robust argument shows that this exponential decay is a generic phenomenon. The average phase can be written as a ratio $Z / Z_Q$ of the physical partition function to the quenched partition function $Z_Q \equiv \int e^{-\Re S}$. The physical partition function, in the large volume limit, behaves thermodynamically as the exponential of the (extrinsic) free energy, and therefore scales as $e^{f V}$, with $f$ the free energy density. The quenched partition function describes some (less interesting) thermodynamic system, and is therefore expected to have the same scaling, but with a different exponent: $e^{f_Q V}$. Thus, the ratio will exponentially decay. This can be avoided only when $f_Q = f$ exactly: when there is no sign problem at all.

The real-time portion of the Schwinger-Keldysh contour gives the action an imaginary part. Note that fields along the real-time portion of the contour do not contribute at all to the real part of the action. Along those directions, importance sampling has no effect and the sign problem is maximally bad --- this is generic to all field theories. Specially to scalar field theory, because the domain of the path integral has infinite measure, the quenched partition function does not converge and the average phase is exactly zero.

\section{Normalizing Flows and Contour Integrals}\label{sec:flows-integrals}
We begin by introducing normalizing flows in the case of a theory with no sign problem. To accelerate the process of sampling from the Boltzmann distribution $e^{-S}$, we can look for a map $\tilde\phi(\phi)$ with the property
\begin{equation}\label{eq:nf-def}
\left(\det \frac{\partial\tilde\phi}{\partial\phi}\right)e^{-S[\tilde\phi(\phi)]} \approx \mathcal N e^{-\phi^2 / 2}
    \text.
\end{equation}
The map $\tilde\phi$ (termed a \emph{normalizing flow\footnote{Strictly speaking, it is the inverse map $\tilde\phi\mapsto \phi$ that is usually referred to as the normalizing flow, as it transforms the distribution $e^{-S}$ into the normal distribution. The convention used here, of working with $\tilde\phi(\phi)$ itself, allows generalization to actions with a sign problem.}} in the machine learning literature) transforms a Gaussian distribution, which can be sampled from efficiently, to the physical distribution desired. The normalization constant $\mathcal N$ is inserted to account for the fact that the partition function $Z$ is generically not equal to the Gaussian integral.

If the flow $\tilde\phi(\phi)$ is exact, it allows expectation values to be computed directly (and is referred to as a trivializing map). A flow which is merely an approximation induces an effective action on the fields $\tilde\phi$ which is unequal to the desired physical action:
\begin{equation}
S_{\mathrm{induced}}(\tilde\phi) = \phi^2 / 2+\log\det \frac{\partial \tilde\phi}{\partial \phi}
    \text,
\end{equation}
where $\phi$ is the preimage of $\tilde\phi$ under the normalizing flow.
To compute the correct expectation values, we must reweight by computing a ratio of expectation values:
\begin{equation}
\langle\mathcal O\rangle
=
\frac{\langle \mathcal O \; e^{S_\mathrm{induced} - S}\rangle_n}{\langle e^{S_\mathrm{induced} - S}\rangle_n}
    \text,
\end{equation}
where $\langle\cdot\rangle_n$ denotes an expectation value with respect to the normal distribution over $\phi$.
In practice, it is often more efficient to use the normalizing flow to generate proposals for a Markov Chain instead --- the distinction will not matter here.

This procedure can begin with any easily sampled distribution. The use of a Gaussian is a convenient choice when the domain of integration is $\mathbb R^N$. For compact domains of integration, a uniform distribution is likely to be a more convenient starting point.

Note also that normalizing flows compose. Given a sequence of distributions $p_1,\ldots,p_k$, and $k-1$ normalizing flows transforming $p_i$ to $p_{i+1}$, the composition of those normalizing flows transforms $p_1$ to $p_k$. This compositional property is preserved by the complex normalizing flows defined below.

This method is clearly not directly applicable to models with a sign problem. The normalizing flow induces an effective action on the physical fields $\tilde\phi$ which is always real\footnote{Or at least, the Boltzmann factor is always real. A noninvertible flow may induce a negative Boltzmann factor.}, and therefore will never match the physical action $S[\tilde\phi]$. We can construct a normalizing flow for the quenched action $\Re S[\tilde\phi]$, but this will at most lead to a polynomial speed up in an exponentially slow algorithm\footnote{Furthermore, as discussed in Sec.~\ref{sec:sk}, the real part of the action for real-time sign problems is typically flat for most directions. Sampling from the quenched action is not hard to begin with.}.

Instead, inspired by the generalized thimble method, we can allow $\tilde\phi(\phi)$ to map real trivial fields $\phi \in \mathbb R^N$ to complex-valued physical fields $\tilde\phi \in \mathbb C^N$. We dub such a construction a \emph{complex normalizing flow}. The condition Eq.~(\ref{eq:nf-def}) remains the same; to guarantee equality of expectation values, we will see that additional constraints on the behavior of $\tilde\phi$ are needed.

Assuming for the moment that Eq.~(\ref{eq:nf-def}) holds exactly, let us see what expectation values are computed.
\begin{equation}\label{eq:nf-exp}
\langle\mathcal O(\tilde\phi)\rangle_n
=
\frac
{\int_{\tilde\phi(\mathbb R^N)}\mathcal D\tilde\phi\;\mathcal O e^{-S}}
{\int_{\tilde\phi(\mathbb R^N)}\mathcal D\tilde\phi\;e^{-S}}
\end{equation}
Although the integrand is the desired one, the domain of integration is incorrect. The physical expectation value $\langle\mathcal O\rangle$ is obtained by an integral over the real plane $\mathbb R^N \subset \mathbb C^N$. The domain of integration used in Eq.~(\ref{eq:nf-exp}) is the image of $\mathbb R^N$ under the map $\tilde\phi$. In order for the two integrals to be guaranteed equal, we must require the following:
\begin{itemize}
    \item The Boltzmann factor $e^{-S[\tilde\phi]}$ is holomorphic, as is the product with the observable $e^{-S}\mathcal O$~\cite{Alexandru:2018ngw}.
    \item The image of $\mathbb R^N$ under $\tilde\phi(\phi)$ is a continuous manifold $\mathcal M\subset \mathbb C^N$.
    \item The contours $\mathbb R^N$ and $\mathcal M$ are connected by a homotopy; that is, there exists a continuous family of manifolds $\mathcal M(t)$ such that $\mathcal M(0) = \mathbb R^N$, $\mathcal M(1) = \mathcal M$, and at no point does $\mathcal M$ pass through a singularity of an integrand.
\end{itemize}
Implicit in the last condition is the requirement that, when the complexified domain is not compact, the asymptotic behavior of the manifold at infinity not change. A change in this asymptotic behavior is considered equivalent to the manifold passing through the singularity at infinity.

From the conditions for equality above, it is clear that a complex normalizing flow induces a manifold of integration $\mathcal M$ of exactly the sort used in the generalized thimble method. For an exactly normalizing flow, the integration along this manifold exhibits no sign problem. Therefore, (exact) complex normalizing flows exist only if there is a manifold which exactly solves the sign problem. In fact, as discussed in Sec.~\ref{sec:flow-existence} below, the converse holds as well: the existence of a manifold with no sign problem implies the existence of an exact normalizing flow.

In cases where the complex normalizing flow is not exact, reweighting is used to recover the precise expectation values as usual. This will generally be necessary throughout the numerical methods explored in this paper.

\section{Existence}\label{sec:existence}
A theory of complex normalizing flows does little good if such flows do not exist for problems of physical interest. The section is devoted to investigating when complex normalizing flows exist. Although in no (non-trivial) case can we show that normalizing flows certainly exist, the evidence suggests that such flows are \emph{more likely} to exist in the case of bosonic (including real-time) sign problems than in the case of fermion sign problems.

First, we construct manifolds that entirely remove \emph{local} phase fluctuations, leaving only global cancellations between different parts of the manifold of integration. This construction uses the holomorphic gradient flow (often used to approximate or define Lefschetz thimbles), defined and characterized in Sec.~\ref{sec:gradient-flow}. The existence of locally sign-free manifolds is argued for in the subsequent section. In Sec.~\ref{sec:global-existence}, we conjecture that locally perfect manifolds behave smoothly as parameters of the action are varied; using this conjecture we argue that perfect manifolds exist for the Schwinger-Keldysh sign problem in scalar field theory. Several examples, where sign-free manifolds can either be found explicitly or shown not to exist at all, are given in Sec.~\ref{sec:examples}; these examples suggest a pattern in which sign-free manifolds generically exist for bosonic, but not fermionic, sign problems. Finally, in Sec.~\ref{sec:flow-existence}, we use well-known results regarding normalizing flows in the real setting to conclude that, conditional on a perfect manifold existing, a complex normalizing flow must exist.

\subsection{Holomorphic Gradient Flow}\label{sec:gradient-flow}
The holomorphic gradient flow is a first-order differential equation used to approximate Lefschetz thimbles~\cite{Alexandru:2015xva,Alexandru:2015sua}. Lefschetz thimbles are the surfaces of steepest descent of $\Re S$ proceeding from critical points of the action. A certain union of the thimbles can be shown to yield the same integral as the real plane $\mathbb R^N$. Because the thimbles are generically sub-optimal in terms of the sign problem, we will ignore them and focus on the behavior of the flow itself. The key result is that, when a Boltzmann factor has local phase fluctuations on the real plane, the holomorphic gradient flow can always be used to find a nearby manifold with an improved sign problem.

The holomorphic gradient flow is defined by
\begin{equation}\label{eq:gradient-flow}
\frac{\d z}{\d t} = \overline{\frac{\partial S}{\partial z}}\text.
\end{equation}
Here the partial derivative $\frac{\partial}{\partial z}$ denotes the usual holomorphic derivative (i.e.\ the Wirtinger derivative). We will assume throughout that the action $S$ is holomorphic in the field variables $z$. This differential equation governs the evolution of a field configuration $z$ through complex space. When applied to all field configurations in a manifold, we obtain a family of manifolds parameterized by the flow time $t$. Note that the flow time is purely fictional, and is unrelated to the physical time (represented as part of the lattice).

Considering the evolution of the field equation, note that the imaginary part of the action never changes, and the real part can only increase (or remain the same, if we begin at a critical point):
\begin{equation}
    \frac{\d S}{\d t} = \left|\frac{\partial S}{\partial z}\right|^2\text.
\end{equation}
This can be taken as a motivation for the holomorphic gradient flow, as by increasing the real part of the action, we may hope to decrease the quenched partition function and improve the average phase. Following this observation, it is convenient to work with the real part of the action $u \equiv \Re S$.

The flow Eq.~(\ref{eq:gradient-flow}) is most frequently applied to manifolds beginning from the real plane ($\mathbb R^N \subset \mathbb C^N$). Let us examine its behavior at early times. Note that, as a consequence of Cauchy's integral theorem, the partition function itself will not be changed by the flow~\cite{Lawrence:2020irw}; only the quenched partition function will change. The change in the quenched partition function is given by
\begin{equation}
\frac{\d}{\d t} Z_Q
=
\frac{\d}{\d t} \int_{\mathbb R^N} \mathcal D x\; e^{-u[z(x)]} \left|\det \frac{\partial z}{\partial x}\right|
\text,
\end{equation}
where we have chosen to parameterize the infinitesimally flowed manifold $z(x)$ by the real plane. Because the domain of integration is unchanged, being the parameterizing real plane for any $t$, we may proceed to inspect the derivative of the integrand.
\begin{equation}\label{eq:derivative-integrand}
    \frac{\d}{\d t} e^{-u} |\det J|
    =
    e^{-u} |\det J| \left[\Re\Tr J^{-1} \frac{\d J}{\d t} - \frac{\d u}{\d t}\right]\text.
\end{equation}
We have already seen that $\frac{\d u}{\d t} = |\frac{\partial S}{\partial z}|^2$ is guaranteed to be non-negative, and positive away from a critical point. The Jacobian term, however, may be larger than the guaranteed-negative term, resulting in a worsening sign problem with flow time $t$. Empirically, this is indeed the case at sufficiently long flow times: the average phase is maximized at some intermediate $t$, rather than in the limit $t\rightarrow \infty$. Beginning from the real plane, however, the Jacobian is the identity, and we find
\begin{equation}
\Tr J^{-1} \frac{\d J}{\d t} = \sum_i \overline{\frac{\partial^2 S}{\partial z_i^2}}\text.
\end{equation}
The real part reduces to $\sum_i \frac{\partial^2 u}{\partial x_i^2}$. Returning to Eq.~(\ref{eq:derivative-integrand}),
\begin{equation}
    \frac{\d}{\d t} e^{-u} |\det J|
    =
    \sum_i e^{-u} \left[ \frac{\partial^2 u}{\partial x_i^2} -
    \left|\frac{\partial S}{\partial z_i}\right|^2
    \right]\text.
\end{equation}
Consider each $i$ independently. Each term is very nearly a total derivative, since
\begin{equation}
    \frac{\partial}{\partial x_i} \left(\frac{\partial u}{\partial x_i}e^{-u}\right)
    = 
    e^{-u} \left[\frac{\partial^2 u}{\partial x_i^2}- \left(\frac{\partial u}{\partial x_i}\right)^2\right]
    \text.
\end{equation}
To connect the two expressions, observe that the magnitude of the derivative of the action can be written
\begin{equation}
    \left|\frac{\partial S}{\partial z_i}\right|^2 =
    \left(\frac{\partial u}{\partial x_i}\right)^2 + \left(\frac{\partial v}{\partial x_i}\right)^2
\end{equation}
where $v \equiv \Im S$.
As a result, we find that the change in the quenched partition function, when starting from the real plane\footnote{The same argument applies to any flat manifold.}, is
\begin{equation}
    \frac{\d Z_Q}{\d t}
    =
    - \int \mathcal D x\;
    e^{- u(x)} \left(\frac{\partial v}{\partial x_i}\right)^2
    \text.
\end{equation}
This is never positive; hence, the sign problem is always improved by a small amount of flow from the real plane. Moreover, as long as the imaginary part of the action is non-constant on the portion of the real plane where $e^{-u}$ is nonvanishing, a small amount of flow will make the quenched partition function strictly smaller.

This is the key result regarding the holomorphic gradient flow: when beginning from the real plane, if $\Im S$ is not constant where $\Re S$ is finite, a small amount of flow is guaranteed to improve the sign problem.

One other property of the holomorphic gradient flow is of interest: regions of $\mathbb C^N$ at which $\Re S$ diverges (becoming large and positive) act as attractors. The flow will collide with these singularities in a finite flow time. This does not cause the evolution of the manifold itself to be ill-defined. The manifold will be continuous, but not smooth, where it intersects singularities of the action. Because the Boltzmann factor vanishes at these singularities, the manifold's behavior there contributes neither to the integral nor to the sign problem.

\subsection{Existence of Locally Perfect Manifolds}\label{sec:manifold-existence}
The holomorphic gradient flow is not guaranteed to result in a perfect manifold at asymptotically large times. The asymptotic manifold under this flow is a union of Lefschetz thimbles. Two features of the Lefschetz thimbles contribute a nonvanishing sign problem. First, on each thimble $\Im S$ is constant, but different thimbles generically have different values of $\Im S$. Thus, cancellations occur between different thimbles, which may become severe when multiple thimbles have similar quenched weights. Second, although $\Im S$ is constant, the phase in the integral is actually $\Im S - \Im \log \det J$, with the Jacobian term coming from the integration measure $\d z$. Each thimble, therefore, comes with local phase fluctuations, which have been found to become severe on large lattices~\cite{Lawrence:2018mve}.

These two problems are sometimes contrasted and referred to as ``global" vs ``local" sign problems. An example of an unremovable global sign problem was given in~\cite{Lawrence:2020irw}: the one-dimensional integral $\int (\cos\theta +\epsilon)\d\theta$, for small $\epsilon$, cannot have its sign problem repaired by any contour deformation. Local sign problems, in contrast, have been found to be removable even where the thimbles fail. In the heavy-dense limit of the Thirring model, numerical experiments show that the flow results in a suboptimal manifold even when a perfect manifold does exist~\cite{Lawrence:2018mve}.

We will see in this section that local sign problems are always removable; that is, a piecewise-smooth contour exists along which there are no locally fluctuating phases, but there may be cancellations between different pieces. Combined with the observation above that at least \emph{some} global sign problems are unremovable, this indicates that the distinction is well defined. Sign problems can be decomposed into a local and global part, with the local part fixable by an appropriate choice of integration contour, but the global part requiring more drastic manipulations.

Now we turn to a procedure, based on the holomorphic gradient flow, for removing local sign problems. Begin with $\mathcal M_0 = \mathbb R^N \subset \mathbb C^N$, and flow for a small amount of time $\epsilon$. This defines a manifold $\mathcal M_1$, parameterized approximately via
\begin{equation}\label{eq:onestep}
    \tilde\phi_1(\phi) = \phi + \epsilon \overline{\frac{\partial S}{\partial \phi}}
    \text.
\end{equation}
(In this discussion, for illustrative purposes, we will expand to linear order in the flow time $\epsilon$, as if a discrete jump was made. To treat the zeros of $e^{-S}$ correctly, it is important to perform a proper evolution by the flow equation instead.) By the argument in the previous section, the quenched partition function on $\mathcal M_1$ is no larger than that on $\mathcal M_0$; moreover, if $\mathcal M_0 \ne \mathcal M_1$, then the quenched partition function is smaller.

Consider the effective action $S_1$ induced by $\tilde\phi_1$; this is a function from $\mathbb R^N$ to the complex numbers. We would now like to flow with respect to this effective action. Doing so would guarantee that the sign problem once again improves. In order for this to make sense, however, $S_1$ must be holomorphic. Naively, it appears not to be. In particular, its definition includes the explicitly antiholomorphic term $\overline{\frac{\partial S}{\partial \phi}}$, preventing us from repeating the last step.

This is a fiction. First consider $\tilde\phi_1(\phi)$ as defined in Eq.~(\ref{eq:onestep}). As written, it appears to contain both a holomorphic and an antiholomorphic term. However, the only aspect of $\tilde\phi_1$ we care about is its definition on $\mathbb R^N$. Functions $\mathbb R^N \rightarrow \mathbb C$ are neither holomorphic nor antiholomorphic; as long as $\tilde\phi_1$ is sufficiently smooth, it can be analytically continued into the complex plane in a purely holomorphic way. Concretely, we can replace Eq.~(\ref{eq:onestep}) by
\begin{equation}\label{eq:onestep-holomorphic}
    \tilde\phi_1(\phi) = \phi + \epsilon \left.\overline{\frac{\partial S}{\partial \phi}}\right|_{\bar \phi}
    \text;
\end{equation}
that is, we evaluate the function $\overline{\frac{\partial S}{\partial \phi}}$ at $\bar \phi$. This parameterizes exactly the same manifold (as the two functions agree on $\mathbb R^N$), but also defines a holomorphic map when evaluated on the rest of the complex plane.

Now we return to the effective action, which after one step is given by
\begin{equation}\label{eq:onestep-effective}
S_1(\phi) = S[\tilde\phi_1(\phi)] -
\log \det \left(
1 +
\epsilon \frac{\partial}{\partial \phi} \left.\overline{\frac{\partial S}{\partial \phi}}\right|_{\bar \phi}
\right)
\text.
\end{equation}
As initially defined, this is an analytic function of the real plane alone. As with $\tilde\phi_1$, we can choose its behavior in the complex plane to be holomorphic, at least in some region around the real plane. We can now flow the real plane again, this time with respect to $S_1$. As before, this is guaranteed to improve the sign problem. We obtain a function $\tilde\phi_2$ which maps the real plane (the domain of $S_1$) to some slightly deformed contour. Composing $\tilde\phi_2 \circ \tilde\phi_1$ yields a map from the domain of the original action $S$ to a deformed contour. Thus, we obtain a new integration manifold $\mathcal M_2 = \tilde\phi_2 \circ \tilde\phi_1(\mathbb R^N)$, which induces an effective action $S_2$, and we repeat.

At every step of this modified flow, we have the freedom to arbitrarily reparameterize the integration manifold. There is no requirement that the parameterizations be ``connected'' from one step to the next. This means that the evolution of the manifold from step to step is non-unique.

What can happen to the manifold in the limit of a large number of steps? As long as the manifold is changing, $Z_Q$ is shrinking; this implies that we cannot reach a cycle. The development of some singular behavior, unremovable by reparameterization, could require us to take ever-smaller steps $\epsilon$. We do not have a formal proof forbidding this; however, numerical experience with the holomorphic gradient flow indicates that it does not create singularities away from zeros of the Boltzmann factor. The last possibility is a fixed point: subsequent manifolds are ever-better approximations (or perhaps equal) to a manifold $\mathcal M_f$ which is unchanged by the holomorphic gradient flow under the effective action.

The properties of such a fixed-point manifold are best understood by considering the corresponding effective action $S_f$. Since $\mathcal M_f$ is unchanged after a step of flow, it must be that the flow vectors $\overline{\frac{\partial S_f}{\partial \phi}}$ lie entirely within the real plane.  Equivalently, since the sign problem is not improved by flow, $\Im S$ must be constant everywhere $e^{-S} > 0$. Critically, this does not imply that $\Im S$ is in fact globally constant, merely that regions of distinct $\Im S$ are separated by vanishing Boltzmann factors.

To summarize: the effective action $S_f$ on the real plane satisfies $e^{-\Re S}\partial \Im S=0$. The imaginary part is locally constant except at places where the entire action diverges (and the Boltzmann factor vanishes). The real plane is thus divided into distinct regions, each with constant $\Im S_f$ and therefore no \emph{local} sign problem, but with the possiblity of cancellations between the regions.

What does this imply about $\mathcal M_f$? Regions on $\mathbb R^N$ where $S_f$ does not diverge correspond to smooth parts of the fixed-point manifold, which have no sign problem when integrated over. These smooth regions terminate where $\mathcal M_f$ intersects with singularites of $S$, either due to a fermion determinant become $0$, or (in the case of bosonic sign problems) where one or more fields $\tilde\phi$ diverge.

The similarities of $\mathcal M_f$ to the Lefschetz thimbles $\mathcal M_T$ are striking. Like $\mathcal M_f$, when the Lefschetz thimbles are parameterized by the real plane, they are separated from each other by regions of vanishing effective Boltzmann factor. Those regions on the real plane correspond to the places in $\mathbb C^N$ where $\mathcal M_T$ intersects with divergences of $S$. The key difference is that, when working with the thimbles, the imaginary part of the physical action is constant but the effective action (due to the Jacobian) may have imaginary fluctuations. The fixed-point manifold $\mathcal M_f$ will have a fluctuating imaginary action, but constant effective action on the real plane.

On the fixed-point manifold, the notion of a ``global'' sign problem becomes clear. Different parts of the manifold have different $\Im S_f$, with those differences apparently not removable by any choice of integration contour. Which physical systems possess global sign problems remains an open question.

Hints of this notion of a ``global'' sign problem are, as mentioned earlier, visible already when considering the Lefschetz thimbles. When the integral over the real line is equal to a sum of integrals over two (or more) thimbles with different phases, it is tempting to disregard the local part of the sign problem on the thimbles, and attribute the cancellations between thimbles to a global sign problem. However, as we will see in Sec.~\ref{sec:examples} below, such a global sign problem on the thimbles does not imply an unremovable global sign problem.

In the context of Lefschetz thimbles, it has been argued that cancellations between different thimbles should not be severe in the infinite volume limit. One such argument proceeds as follows~\cite{cohen1}. Each thimble is associated to a critical point of the action; i.e., a classical solution to the equations of motion. At large volumes, we may reasonably expect the integral to be dominated by thimbles associated with large-scale classical solutions. These solutions, and therefore their associated thimbles, persist as we enlarge the volume. Therefore, we may now talk about ``a thimble'' across multiple volumes. Each thimble's contribution to the path integral should grow thermodynamically, defining a per-thimble free energy. Unless protected by some symmetry, each of these free energies will generically be different, causing one thimble to dominate in the large-volume limit. Even in the case of the thimbles, this argument is not a proof. Its applicability to the fixed-point manifolds is particularly unclear.

\subsection{Existence of Perfect Manifolds}\label{sec:global-existence}

In the restricted case of \emph{polynomial} actions (this excludes lattice models with a fermion determinant), we can show that globally perfect manifolds are likely to exist, provided that locally perfect manifolds depend smoothly on the parameters of the action. To be precise, the conjecture we need is:
\begin{conjecture*}
Let $S_t$ be a continuous family of actions, and let $\mathcal M$ be a manifold on which $e^{-S_0} \d z$ has no local phase fluctuations. Then there exists a continuous family of manifolds $\mathcal M_t$, with $\mathcal M_0 = \mathcal M$, such that $e^{-S_t} \d z$ has no local phase fluctuations on $\mathcal M_t$.
\end{conjecture*}
This conjecture empirically holds in several one-dimensional models explored in the next section. It is also motivated by thinking of normalizing flows as analytic functions not just of the field variables, but also of the parameters of the action.

Suppose we start from an action $S_0$ that has no sign problem on the real plane, whether local or global. Later actions $S_t$ have a sign problem. By definition the corresponding manifolds $\mathcal M_t$ have no local sign problem; can a global sign problem be created?

One way for a global sign problem to be created, without requiring discontinuous behavior of the family $\mathcal M_t$, is for singularities of the action to intersect with the manifold. In the case of the Schwinger-Keldysh action Eq.~(\ref{eq:sk}) and other polynomial actions, however, there are no singularities of the action except at infinity. Any global sign problem must come from regions of $\mathcal M_t$ on which all fields can become arbitrarily large.

Creating a global sign problem, therefore, implies introducing a new such region of $\mathcal M_t$. This is possible\footnote{A previous version of this paper claimed that the creation of such a region constituted a discontinuous operation on the family of manifolds. This is not true, as long as this region of $\mathcal M_t$ can be deformed away from infinity while keeping the action bounded below.}, but much more difficult than intersecting a zero which lies a finite distance from the origin. If we are to hold to the previous conjecture, this argument suggests that this family of manifolds is not only locally perfect, but in fact globally perfect.

\subsection{Examples}\label{sec:examples}
\begin{figure*}
    \centering
    \includegraphics[width=0.43\linewidth]{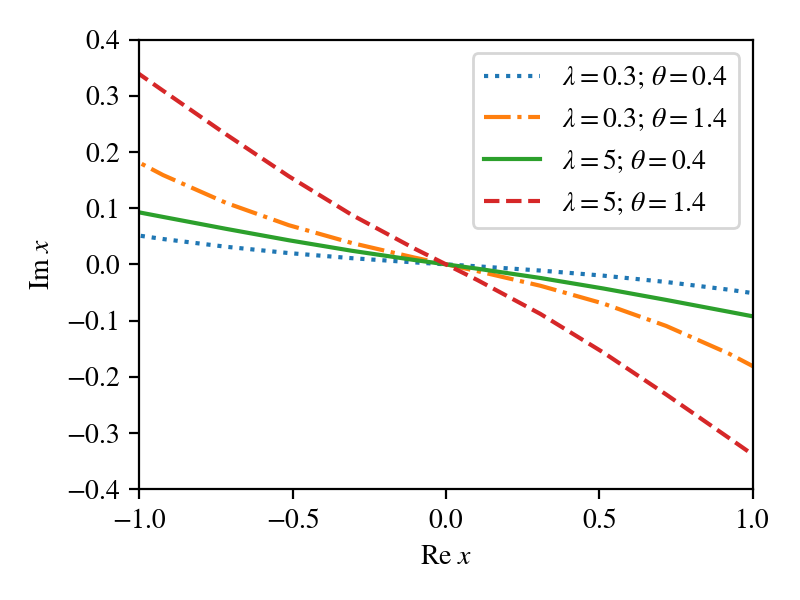}\hspace{0.05\linewidth}
    \includegraphics[width=0.43\linewidth]{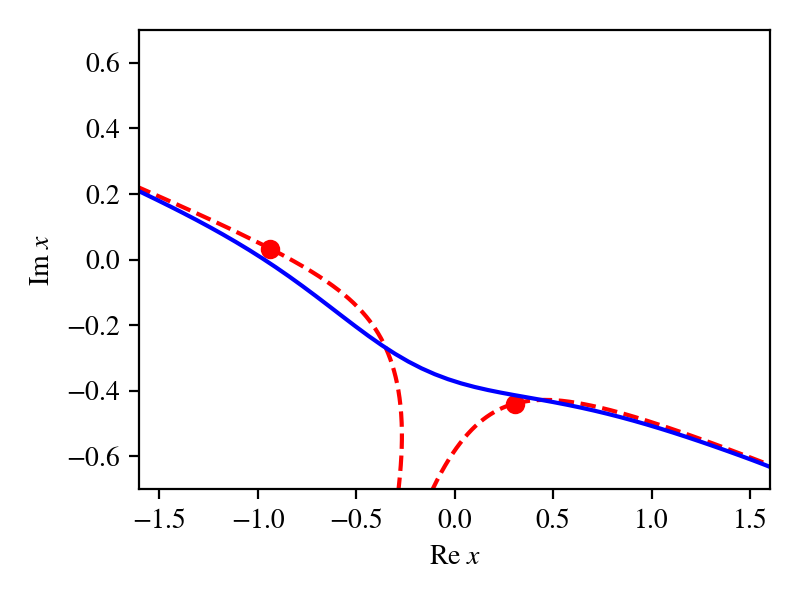}
    \caption{Perfect manifolds, found by numerical search, for the one-dimensional integral defined by Eq.~(\ref{eq:example-action}). All manifolds have an average sign measured to be within $10^{-5}$ of unity. The left panel shows the case of $m^2 = 1$, $b=0$; the right panel shows a single example with $m^2 = -1$, $b=\lambda=\theta=1$, contrasted with the Lefschetz thimbles (red dashed lines).}
    \label{fig:perfect-examples}
\end{figure*}

In one dimension, manifolds with no sign problem can be readily found by a numerical search. As an example, the left panel of Fig.~\ref{fig:perfect-examples} shows such manifolds for the action
\begin{equation}\label{eq:example-action}
    S = m^2 x^2 + \lambda e^{i \phi}x^4 + i b x^3
\end{equation}
for various values of $m^2$, $\lambda$, $b$, and $\phi$. These examples motivate the conjecture that similarly structured sign problems (in particular, those with a polynomial action) generally admit perfect manifolds.

The right-hand panel of Fig.~\ref{fig:perfect-examples} demonstrates the existence of a perfect manifold in a situation where the Lefschetz thimbles display both global and local cancellations. There are two thimbles, each of which has curvature and therefore a local sign problem coming from the nontrivial Jacobian. The imaginary parts of the action at the two contributing critical points are different, meaning that the two thimbles also exhibit cancellations between each other, worsening the sign problem. Nevertheless, a single manifold exist which has neither sort of sign problem.

The availability of perfect manifolds does not hold even for all one-dimensional integrals, however. A simple example, not physically motivated, was given in~\cite{Lawrence:2020irw}. The integral of $(\cos \theta + \epsilon)$ has a sign problem of order $\epsilon^{-1}$ for small $\epsilon$. For sufficiently small $\epsilon$, it cannot have its sign problem removed by any contour deformation. This is readily confirmed by noticing that the magnitude of $\cos(a+ib)$ (which is the quenched Boltzmann factor) is minimized when $b=0$. The integral along the real line will then have the smallest quenched partition function, and therefore the best possible sign problem.

In the previous section, we discussed how a global sign problem could be created when singularities of the action intersected with a locally perfect manifold. The case of $(\cos \theta + \epsilon)$ is a clear demonstration of this phenomenon. At $\epsilon > 1$, there are two zeros of the Boltzmann factor, at $\Re \theta = 0$ and $\Im\theta = \pm \cosh^{-1} \epsilon$. As epsilon is lowered, these move towards the real line; at $\epsilon=1$ they merge at $\theta = 0$. At this point, no global sign problem yet exists, but the locally perfect manifold now passes through a zero. Continue lowering $\epsilon$, and the two zeros again split, now at $\Re \theta = \pm \cos^{-1}\epsilon$. Although the manifold has never changed, it now consists of segments with cancelling phases.

Let us now consider a more physical model: the $0+1$-dimensional Thirring model as studied in~\cite{Alexandru:2015sua,Alexandru:2015xva}. The Boltzmann factor defining this model is
\begin{widetext}
\begin{equation}
    e^{-S} = 
    \exp\left(\frac{1}{2 g^2} \sum_i \cos z_i\right)
    \det_{i,j}\left[
    m \delta_{i,j} + \frac 1 2 \left(
    e^{\mu + i z_i} \delta_{i+1,j}
    - e^{-\mu - i z_{j}} \delta_{i-1,j}
    + e^{-\mu-i z_j} \delta_{i,1} \delta_{j,\beta}
    - e^{\mu + i z_i} \delta_{j,1} \delta_{i,\beta}
    \right)
    \right]
    \text,
\end{equation}
\end{widetext}
where $g$ is a coupling constant, $\mu$ is the chemical potential (and origin of the sign problem), and $m$ is the bare mass. The $z_1,\ldots,z_\beta$ are the degrees of freedom being integrated over; there are $\beta$ links on the lattice.

The fermion determinant depends only on the sum of the fields $\beta \sigma = \sum_i z_i$. A natural simplification, therefore, is to consider the ``mean-field'' model, a one-dimensional integral with Boltzmann factor
\begin{equation}
    e^{-S(\sigma)}
    =
    e^{\frac{\beta}{2 g^2} \cos\sigma}
    \big[
    \cos(\beta (\sigma - i\mu)) + 1
    \big]
    \text.
\end{equation}
We have taken $m=0$ for convenience (and neglected an overall normalization). 

For our purposes, an interesting limit is that of large $\beta$, while keeping the coupling and chemical potential both of order unity. Numerical experiments indicate that there is no contour that exactly solves the sign problem with these parameters --- indeed, the average phase falls exponentially in $\beta$, as one would expect. This holds even if we neglect the Jacobian. In particular, write the quenched Boltzmann factor explicitly in terms of the real and imaginary parts of $\sigma = \sigma_R + i \sigma_I$:
\begin{multline}
|e^{-S}| = 
    e^{\frac{\beta}{2 g^2} \cos\sigma_R \cosh \sigma_I}
    \big|1 + \cos\beta\sigma_R \cosh(\beta(\sigma_I - \mu)) \\- i \sin\beta\sigma_R \sinh(\beta(\sigma_I-\mu))
    \big|
    \text.
\end{multline}
The quenched partition function can be given a lower bound by minimizing $\int |e^{-S[\sigma_R,\sigma_I(\sigma_R)]}|$ over all functions $\sigma_I$. The minimization over $\sigma_I$ can be done individually for each $\sigma_R$. Even this lower bound on the quenched partition function still falls exponentially above the physical partition function.

Note that the fact that no manifold exists to resolve the mean-field sign problem does not prove that no manifold exists that resolves the sign problem of the original theory: it is merely suggestive. It seems plausible that a similar technique could be used to establish the impossibility of the original sign problem.

This fermionic example differs sharply from the Schwinger-Keldysh action (and from Eq.~(\ref{eq:example-action})). The action is not a polynomial, and relatedly, the Boltzmann factor falls to zero away from infinity (and nearly on the real plane). If the failure to have a perfect manifold is related to these features, then we expect fermionic sign problems to frequently be unresolvable via contour deformation, while bosonic sign problems would generically be resolvable.

\subsection{Existence of Normalizing Flows}\label{sec:flow-existence}
A parameterization $\tilde\phi(\phi)$ of a manifold with no sign problem induces an effective action on the real plane that is always real. Thanks to the composability of normalizing flows, the problem of finding a complex normalizing flow reduces to the problem of finding an ordinary normalizing flow for that effective action. Provided that this can be done, the existence of a perfect manifold implies the existence of a complex normalizing flow.

As it happens, given probability distributions $p(x)$ and $\pi(\tilde x)$, a map $x\rightarrow\tilde x$ always exists such that the measure $p(x) \d x$ induces the measure $\pi(\tilde x) \d \tilde x$; that is, such that
\begin{equation}\label{eq:cdf1}
    p(x) \left(\det \frac{\partial \tilde x}{\partial x}\right)= \pi[\tilde x(x)]
    \text.
\end{equation}
The construction is simplest, and unique, in one dimension. Define the cumulative distribution functions $P$ and $\Pi$, of $p$ and $\pi$ respectively:
\begin{equation}\label{eq:cdf2}
    P(x) = \int_{-\infty}^{x} \d x' \; p(x)
    \text.
\end{equation}
The CDF can be seen as a normalizing flow from a probability distribution to the uniform distribution on the unit interval. Therefore, the desired map is given by $\Pi^{-1} \circ P$.

In the multidimensional case such maps are known to exist as well, but cease to be unique. Finding maps with desirable properties is an active area of research; see~\cite{villani2003topics} for a review.

Heuristically, one expects a normalizing flow to depend smoothly --- even analytically --- on the parameters of the action. This is certainly the case for the one-dimensional models considered above. This provides a new perspective on the conjecture of Sec.~\ref{sec:global-existence} above. Suppose the action $S_\lambda(z)$ depends on a potentially complex parameter $\lambda$, but is real (having no sign problem) when $\lambda$ is real. A smooth family of normalizing flows $z = \phi(x;\lambda)$ for real $\lambda$ can be analytically continued to complex $\lambda$, defining a smoothly varying family of locally perfect manifolds as per the conjecture.

The remainder of this work is dedicated to the task of finding approximate normalizing flows, under the assumption that such flows exist.

\section{Perturbing Flows}\label{sec:perturbation}
In principle, a complex normalizing flow can be trained in much the same way as a regular normalizing flow. In practice, this training is a difficult process. One principal reason, closely linked to the sign problem, is that when comparing Boltzmann factors, a difference of $2\pi$ in the action is invisible. As a result, if the physical Boltzmann factor is $1$ and the induced Boltzmann factor is $-1$, the gradient descent procedure has no way to know whether the induced $\Im S$ should be changed by $\pi$ or $-\pi$ (or perhaps $3\pi$). Circumventing this requires either maintaining a normalizing flow which is always ``within $\pi$'' of being exact, or defining the flow in such a way that the imaginary part of the induced action is itself well defined. Instead, we will work in the spirit of~\cite{Lawrence:2020kyw}, and construct normalizing flows in perturbation theory.

\subsection{Leading Order}
A normalizing flow need not begin with a Gaussian distribution. In the general case, the condition for a normalizing flow reads
\begin{equation}\label{eq:nf-general}
\left(\det \frac{\partial\tilde\phi}{\partial\phi}\right)e^{-S[\tilde\phi(\phi)]} = \mathcal N e^{-S_0(\phi)}
    \text,
\end{equation}
where $S_0$ is the action defining the original probability distribution, which $\phi\mapsto\tilde\phi$ transforms into $e^{-S}$.
Consider the case where $S$ is merely a perturbation of $S_0$; that is, where
\begin{equation}
    S = S_0 + \lambda \mathcal O
\end{equation}
for small $\lambda$. When $\lambda = 0$, a suitable normalizing flow is simply $\tilde\phi = \phi$. For small $\lambda$, we expand $\tilde\phi$ as a power series: $\tilde\phi = \phi + \lambda \Delta^{(1)}$. Returning to Eq.~(\ref{eq:nf-general}) and expanding to leading order in $\lambda$, we find the differential equation for $\Delta^{(1)}$:
\begin{equation}\label{eq:p1}
\nabla\cdot\Delta^{(1)}
- \Delta^{(1)} \cdot \nabla S_0
= 
\mathcal O-\langle\mathcal O\rangle
\text.
\end{equation}
Here the expectation value $\langle \mathcal O \rangle$ is evaluated with respect to the original action $S_0$.
We can obtain Eq.~(\ref{eq:p1}) more quickly simply by considering the integral of $\nabla\cdot\left(\Delta e^{-S_0}\right)$, for any $\Delta$ that decays at infinity, or diverges sufficiently slowly. As the integral of a total derivative, it must vanish. This implies that the expectation value of $\nabla\cdot\Delta - \Delta\cdot \nabla S_0$ vanishes as well.

When perturbing from a free theory (that is, when $S_0$ defines a Gaussian), Eq.~(\ref{eq:p1}) can be solved exactly. In particular, with
\begin{equation}\label{eq:pf1-action}
    S = \sum_{ij} \phi_i M_{ij} \phi_j + \lambda \sum_i \Lambda_i \phi_i^4\text,
\end{equation}
the perturbative flow $\Delta^{(1)}$ is given by
\begin{equation}\label{eq:pf1-sk}
    \Delta^{(1)}_i = -\sum_{j} \left[
    \frac 1 2 M^{-1}_{ij} \Lambda_j \phi^3_j +
    \frac 3 4 M^{-1}_{ij} M^{-1}_{jj} \Lambda_j \phi_j
    \right]
    \text.
\end{equation}
Note that Eq.~(\ref{eq:pf1-action}) is a generalization of the Schwinger-Keldysh action Eq.~(\ref{eq:sk}). Any two Gaussians are trivially connected by a normalizing flow, and as noted earlier, normalizing flows compose. Thus, Eq.~(\ref{eq:pf1-sk}) implicitly defines a (perturbative) normalizing flow for the Schwinger-Keldysh sign problem in $\phi^4$ field theory.

Unfortunately, Eq.~(\ref{eq:nf-general}) is not the only condition constraining a \emph{complex} normalizing flow. As discussed in Sec.~\ref{sec:flows-integrals}, the asymptotic behavior of the contour $\tilde\phi(\mathbb R^N)$ must match that of the real plane; i.e., the two manifolds must be in the same homology class. It is not a surprise that the perturbative flow Eq.~(\ref{eq:pf1-sk}) violates this condition, as the perturbative expansion is equivalent to an expansion in small fields $\phi$, while the asymptotic behavior is purely determined by the behavior of $\Delta^{(1)}$ when $\phi$ is large.

\begin{figure*}
    \centering
    \includegraphics[width=0.35\linewidth]{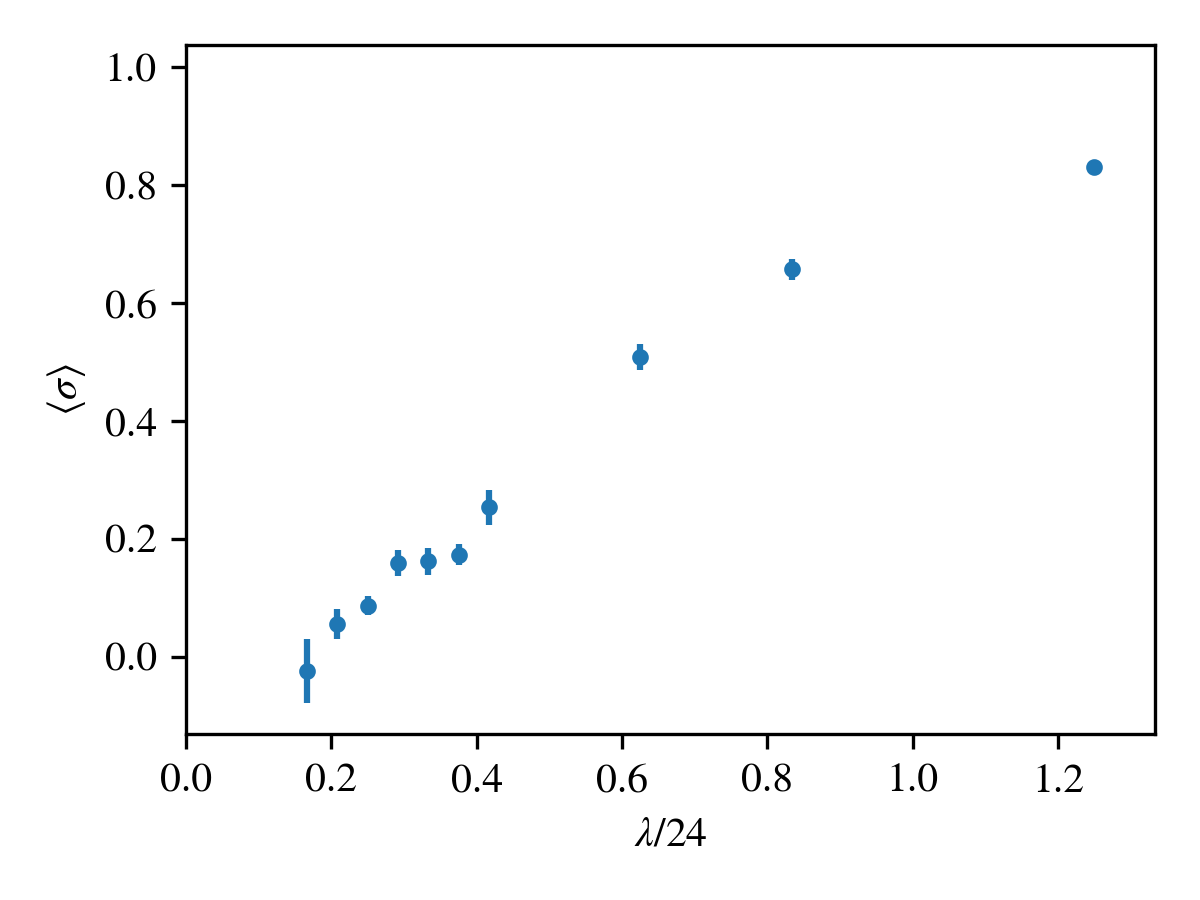}
    \hspace{0.1\linewidth}
    \includegraphics[width=0.35\linewidth]{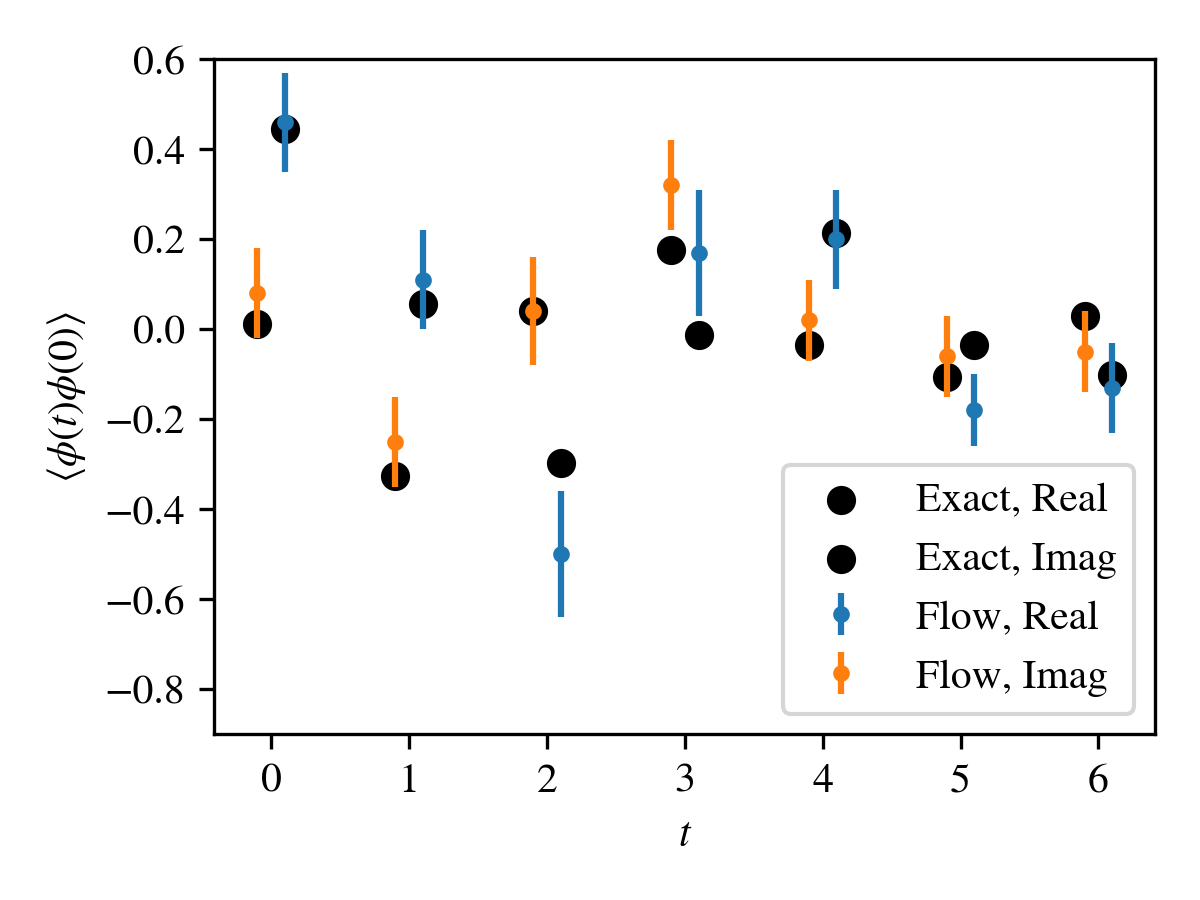}
    \caption{Simulations with the normalizing flow computed in the strong-coupling expansion to leading order. On the left, the resulting sign problem is computed on a lattice with $N_\beta = 2$, $n_t = 5$, $m=0.5$, as a function of the coupling $\frac{\lambda}{24}$. The real-time correlator $\langle \phi(t)\phi(0)\rangle$ is shown on the right, at the same temperature and with $m=0.5$ and $\frac{\lambda}{24} = 0.33$. The solid lines labelled `Exact' include the same Trotterization errors present on the Schwinger-Keldysh lattice.}
    \label{fig:strong}
\end{figure*}

To get the correct asymptotic behavior, we can work instead in the strong coupling expansion to obtain $\Delta^{(1,\mathrm{strong})}$, which becomes a good approximation at large $\phi$. For an action of the form of Eq.~(\ref{eq:pf1-action}), it is convenient to construct our normalizing flow as a sequence of four maps:
\begin{enumerate}
    \item Map the distribution $e^{-\frac{\phi^2}{2}}$ to $e^{-\psi_1^4}$ via $\psi_1=F_1(\phi)$.
    \item Rotate and scale the complex plane via $\psi_2 = F_2(\psi_1)$ to obtain the distribution $e^{-\Lambda \psi_2^4}$.
    \item Introduce a perturbative quadratic piece via a perturbative flow $\psi_3 = F_3(\psi_2) = \psi_2 + \frac 1 {\sqrt{\lambda}}\delta^{(1)}(\psi_2)$. The resulting distribution is $e^{-S'(\psi_3)}$, where
    \begin{equation}\label{eq:rescaledS}
        S'(\psi) = \sum_i\Lambda_i \psi_{i}^4 + \frac{1}{\sqrt{\lambda}}\sum_{ij}\psi_{i} M_{ij}\psi_{j}
        \text.
    \end{equation}
    \item Rescale the fields to restore the correct field normalization via $\tilde\phi = F_4(\psi_3)$, finally obtaining the desired distribution $e^{-S(\tilde\phi)}$, with the action defined in Eq.~(\ref{eq:pf1-action}) 
\end{enumerate}
Note that the first two maps,  $F_1$ and $F_2$, factor into one-dimensional maps, which can be obtained straightforwardly via the prescription following Eq.~(\ref{eq:cdf2}). Accordingly, $F_1$ can be written as
\begin{eqnarray}
    F_1(\phi) &=& \Pi^{-1} \circ P\text{, with}\\
    \Pi(\phi) &=& \frac 1 2 + \frac 1 2 \left( 1 - \frac{\Gamma\left[1/4,\phi^4\right]}{\Gamma(1/4)} \right)\sgn \phi\\
    P(\phi) &=& \frac{1}{2}\left( 1+\Erf(\phi/\sqrt{2}) \right)
    \text.
\end{eqnarray}
Above, $\Gamma(x)$ is the gamma function, $\Gamma(s,x)$ is the upper incomplete gamma function, and $\Erf(x)$ is the error function.

The second map is given by simply multiplying $\phi$ by $\Lambda_i^{-1/4}$. This is a rotation of the complex plane on most of the lattice, with an additional scaling factor of $2^{1/8}$ on the corners of the Schwinger-Keldysh contour. Thus the map $F_2$ is defined as 
\begin{equation}
    F_2(\phi) = \phi/\Lambda_i^{1/4}
    \text.
\end{equation}

The map $F_3$ is where the strong coupling expansion is performed. The differential equation for $\delta^{(1)}$ is of the form of Eq.~(\ref{eq:p1}), with $\mathcal O = \sum_{ij}\phi_i M_{ij}\phi_j$. The expectation value $\langle \mathcal O\rangle$ must now be evaluated with respect to the leading-order action $S_0=\sum_{i}\Lambda_{i}\phi_i^4$. Expressed in terms of $f_i = \delta^{(1)}_i(\phi) e^{-\Lambda_i\phi_i^4}$, and using the fact that $\langle \phi_i \phi_j\rangle$ vanishes when $i \ne j$ (in the strong coupling limit), the differential equation reads
\begin{equation}
    \frac{\partial f_i}{\partial \phi_i} e^{\Lambda_i \phi_i^4} -\sum_{j} M_{ij}\phi_{i}\phi_{j} = - M_{ii} \langle \phi_i^2 \rangle
    \text.
\end{equation}
The expectation value required is $\langle \phi_i^2\rangle = \frac{\Gamma(3/4)}{4\Gamma(5/4) \sqrt{\Lambda_i}} $.
Using the fact that only diagonal and nearest-neighbor terms of $M$ are non-zero, the solution is
\begin{widetext}
\begin{equation}\label{eq:nf-strong}
    \delta^{(1)}_{i}(\phi) = e^{\Lambda_i \phi_i^4}M_{ii}\left[-\frac{ \phi_i^3 \Gamma[\frac{3}{4},\Lambda_i \phi_i^{4}]}{4(\Lambda_i\phi_i^{4})^{3/4}}+ \frac{\langle \phi_i^{2} \rangle \phi_i \Gamma[\frac{1}{4},\Lambda_i \phi_i^4]}{4(\Lambda_i \phi_i^{4})^{1/4}}\right]
    + \sum_{j \in \{i-1,i+1\}}
    e^{\Lambda_i\phi_i^4}\frac{\sqrt{\pi}}{4 \sqrt{\Lambda_i}}\left[ \Erf(\sqrt{\Lambda_i}\phi_i^{2}) - C\right] M_{ij} \phi_{j}
    \text.
\end{equation}
\end{widetext}
Above, $(\cdot)^{1/4}$ refers specifically to the principle fourth root. A specific choice of $C = 1$ gives a solution which vanishes at $\psi_i \rightarrow \infty$ and is oscillation-free. 
 
 Finally, $F_4$ rescales the field by a factor of $\lambda^{1/4}$:
 \begin{equation}
     F_4(\phi) = \phi/\lambda^{1/4}\text.
 \end{equation}
Putting it all together, the entire perturbative flow from $e^{-\sum_i \psi_i^2}$ to $e^{-S(\phi)}$ at the leading order is
 \begin{equation}
 \phi + \Delta^{(1,\mathrm{strong})}(\phi) =
 \left[F_4\circ F_3\circ F_2\circ F_1
 \right](\phi)
 \text.
 \end{equation}

The left panel of Fig.~\ref{fig:strong} shows the average phase obtained by this flow on a $12$-site lattice with $m=0.5$, as the coupling is varied. As expected, at strong coupling, the sign problem is almost entirely removed, whereas at sufficiently small coupling the average phase is too small to be distinguished from zero. As a check of the correctness and convergence of the flow, the right panel of the same figure shows the real-time correlator obtained with $m = 0.5$ and  $\frac{\lambda}{24} = 0.33$, compared with an exact Hamiltonian calculation. The lattice behind this calculation has $14$ sites (two thermal links and six temporal links in each direction), and the average phase was computed to be $\langle\sigma\rangle = 0.096(5)$.

As with many other methods for mitigating the sign problem, the parameters in an ansatz flow can be tuned nonperturbatively ~\cite{Alexandru:2018fqp,Wynen:2020uzx,Mori:2017nwj,Mori:2017pne,Lawrence:2020kyw}. One method for achieving this is similar to the standard technique for training real normalizing flows~\cite{Albergo:2019eim,Kanwar:2020xzo,Boyda:2020hsi}: the normal distribution is sampled from and Eq.~(\ref{eq:nf-def}) enforced on the samples via gradient descent.

\subsection{Extracting Expectation Values}\label{sec:direct-expect}
When the action $S_0$ defines a probability distribution from which sampling can be performed efficiently, Eq.~(\ref{eq:p1}) provides a means to approximately sample from the perturbed distribution defined by $S$. However, that equation is still valid when $S_0$ is itself hard to sample from (and perhaps afflicted with a sign problem). Any vector field $\Delta_i$ corresponds to some observable $\mathcal O$ whose expectation value is known --- and for a fixed desired observable, a numerical solution to $\Delta$ can be attempted, which will automatically yield the expectation value.

The differential equation is in a number of dimensions equal to the number of sites on the lattice. Neural networks have been profitably applied to solving such high-dimensional differential equations~\cite{han2018solving}. Our strategy is as follows. A multi-layer perceptron (MLP), with parameters labelled $W$, will represent $\Delta$ as a function of the (real) fields $\phi$; a single additional training parameter $E$ represents $\langle \mathcal O\rangle$. We train these parameters with respect to the cost function
\begin{multline}\label{eq:mlp-cost}
    C(W,E) = \int \d \phi\;e^{-\phi^2 / 2}\times\\
    \left| \nabla \cdot \Delta_W(\phi)
    -
    \Delta_W(\phi)\cdot \nabla S(\phi)
    - E + \mathcal O(\phi)
    \right|^2
    \text,
\end{multline}
which is estimated by randomly sampling from the Gaussian distribution in $\phi$.

\begin{figure*}
    \centering
    \includegraphics[width=0.35\linewidth]{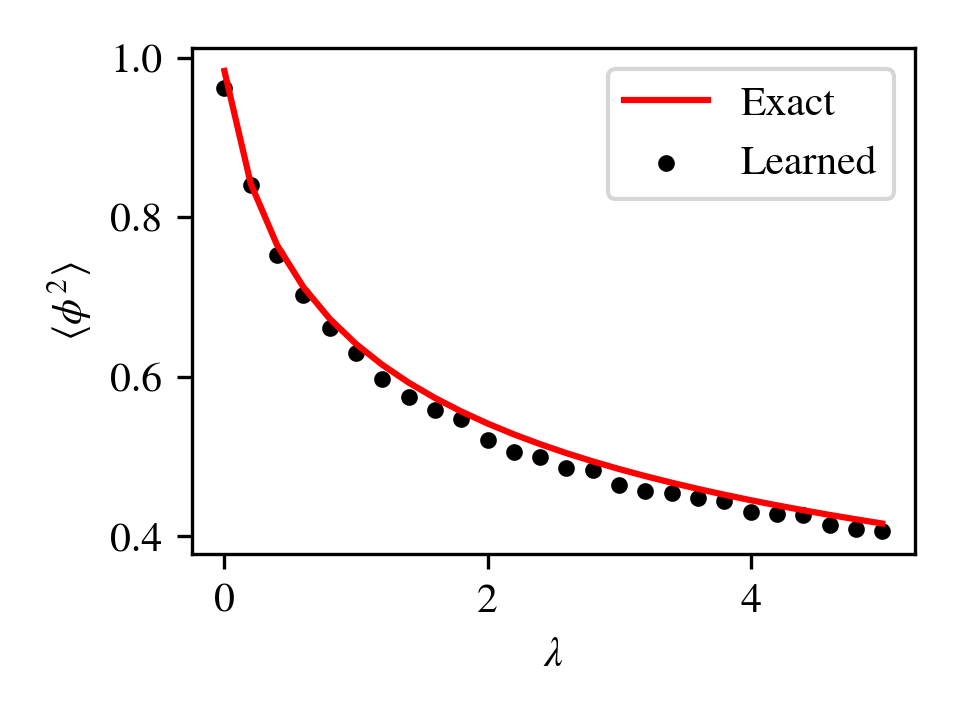}
    \hspace{0.05\linewidth}
    \includegraphics[width=0.45\linewidth]{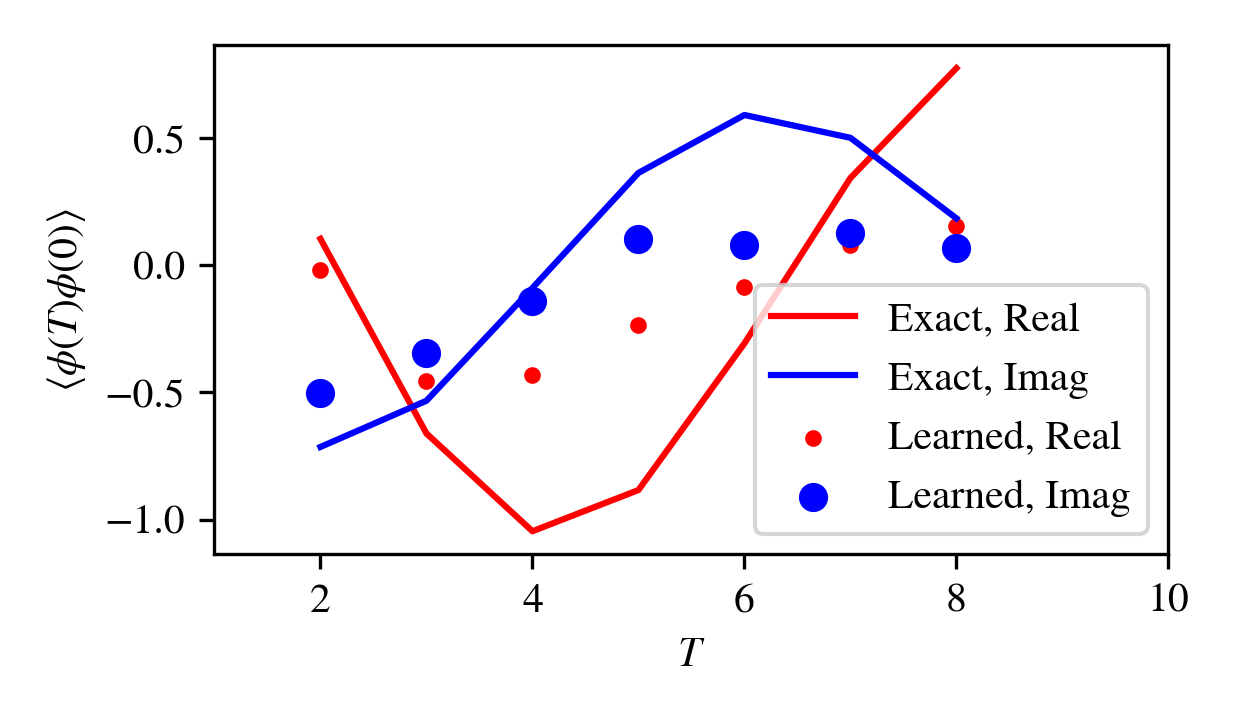}
    \caption{Evaluation of expectation values via the machine learning method of Sec.~\ref{sec:direct-expect}. On the left, a $10$-site lattice with no real-time evolution, with $m=0.5$ and varying the coupling. On the left, lattices of varying real-time extent with $\beta = 2$, $m=0.5$, and $\lambda = 0.5$. In both cases, exact results are shown by the solid line.}
    \label{fig:direct-ev}
\end{figure*}

The left panel of Fig.~\ref{fig:direct-ev} showcases this method on $0+1$-dimensional scalar field theory --- equivalent to the model of Eq.~(\ref{eq:sk}) with no real-time evolution. The lattice parameters are $m=0.5$, $\beta = 10$; expectation values are given as a function of the coupling. A two-layer MLP is used with hyperbolic tangent as an activation function. The method is seen to have reasonable agreement with the exact answer across all couplings.

This method has serious drawbacks. Most importantly, it represents an uncontrolled approximation. The error in the estimation of $\mathcal \langle\mathcal O\rangle$ is due to the shortcomings of the ansatz used for $\Delta$, rather than insufficient statistics; therefore, this error cannot be estimated with bootstrap nor removed with a larger number of samples.

In the case of the Schwinger-Keldysh action with $N_t > 0$, another issue emerges. The most natural cost function for training $\Delta$ would be
\begin{multline}
    C'(W,E) = \int \d\phi\;e^{-S}\times\\
    \left| \nabla \cdot \Delta_W(\phi)
    -
    \Delta_W(\phi)\cdot \nabla S(\phi)
    - E + \mathcal O(\phi)
    \right|^2
    \text.
\end{multline}
The task of evaluating this cost function itself has a sign problem; estimating its derivatives with respect to the MLP parameters has a related signal-to-noise problem. Training with respect to the cost function of Eq.~(\ref{eq:mlp-cost}) has no such difficulty, but a small value of that cost function does not imply that $\Delta$ is a good approximation in terms of the `true' cost function. This mismatch allows an apparently good fit to correspond to very inaccurate expectation values.

The right panel of Fig.~\ref{fig:direct-ev} showcases this failure more clearly, in the case of $N_\beta = 2$, for various real-time evolutions $N_t$. Shown is the expectation value $\langle \phi(t) \phi(0)\rangle$ for several time separations $t$, at an inverse temperature of $\beta = 2$, with lattice parameters $m=0.5$ and $g = 0.5$. Although some qualitative features of the true correlator are reproduced, for larger time evolutions, the learned correlator systematically diverges from the exact answer.

\section{Further Discussion}\label{sec:discussion}
This paper introduced the notion of complex normalizing flows, which extend the applicability of normalizing flow-based methods to (some) models afflicted with a sign problem. Unlike real normalizing flows, which always exist, complex normalizing flows exist only when an integration contour is available which exactly removes the sign problem. Approximate normalizing flows can be constructed in perturbation theory, but low-order perturbative approximations were found to not have a tractable sign problem on lattices with more than $\sim 20$ sites. Thus, at the order computed in this paper, perturbatively constructed normalizing flows do not perform as well as methods based on the holomorphic gradient flow~\cite{Alexandru:2016gsd,Alexandru:2017lqr}. Moreover, given the exponential cost associated with going to higher orders, this particular construction of normalizing flows is unlikely to represent a practical attack on sign problems of higher-dimensional theories.

The general question of when sign problem-solving integration contours exist remains open. For simple fermionic models, they can be readily shown not to exist, although the possibility remains that the manner of integrating out fermions could be changed to remedy this. Simple models inspired by the Schwinger-Keldysh action \emph{do} have contours which exactly solve the sign problem. Conjecturally, this is a generic feature of polynomial actions.

The Schwinger-Keldysh sign problem for scalar fields comes from a polynomial action, but this is far from an unusual property. Consider the case of $SU(N)$ gauge theories in the absence of fermions. The complexification of $SU(N)$ is the group of complex $N\times N$ matrices $U$ obeying $\det U = 1$, $SL(N;\mathbb C)$. On this space, the standard Wilson action and its many improvements can be written holomorphically as a polynomial of $U$ and $U^{\dagger}$. Cramer's rule for the inversion of matrices provides an expression for $U^{-1}$ in terms of the elements of $U$, in which the only non-polynomial factor is $(\det U)^{-1}$. In the case of $SL(N;\mathbb C)$ matrices, this factor is always $1$ and can be neglected. What remains in the action is a polynomial of the field variables.

We argued that contours always exist which \emph{locally} solve the sign problem, by describing an iterative procedure for removing local fluctuations in $\Im S$, and studying the properties of its fixed points. Analogous to the Lefschetz thimbles\footnote{In fact, the regions on the real plane of constant $\Im S_f$ are thimbles of the fixed-point effective action.}, a fixed-point manifold of this procedure can be decomposed into several smooth pieces, separated from each other by singularities of the physical action. The key difference is that on this manifold, the imaginary part of the effective action is constant, rather than the imaginary part of the physical action.

If the evidence presented earlier is taken at face value, it is likely that manifolds exist that resolve the real-time bosonic sign problem. It is critical to note that this is not equivalent to a \emph{solution} to that sign problem. Difficulties could still exist with the computational task of finding such manifolds, or there may be no efficient algorithms for sampling from them. In fact, in the case of a real-time sign problem with arbitrary time-dependent source terms (which is still a polynomial action), this is the expected result: it was shown in~\cite{jordan2018bqp} that the task of computing amplitudes in such a context is BQP-hard.

\begin{acknowledgements}
We are indebted to Andrei Alexandru, Tom Cohen, Frederic Koehler, Henry Lamm, and Michael Wagman for many useful discussions. We are additionally grateful to Henry Lamm for reviewing a previous version of this manuscript. S.L.~is supported by the U.S.~Department of Energy under Contract No.~DE-SC0017905. Y.Y.~is supported by the U.S.~Department of Energy under Contract No.~DE-FG02-93ER-40762 and by the Jefferson Science Associates 2020-2021 graduate fellowship program.
\end{acknowledgements}

\bibliography{nf}

\end{document}